%% file: main.tex
\documentclass{iopjournal}

\graphicspath{{../}}

\usepackage{amsmath,amssymb}
\usepackage{bm}

\usepackage{tikz}
\usetikzlibrary{positioning, calc, arrows.meta}

\usepackage{natbib}
\usepackage{aas_macros}
\defcitealias{death_star}{PS21}

\newcommand{\Msun}{M_\odot}

\newcommand{\Myr}{\mathrm{Myr}}
\newcommand{\Gyr}{\mathrm{Gyr}}
\newcommand{\pc}{\mathrm{pc}}
\newcommand{\au}{\mathrm{au}}
\newcommand{\kms}{\mathrm{km\,s^{-1}}}

\newcommand{\tL}{t_\mathrm{L}}

\begin{document}

\articletype{Paper} 

\title{Formation of stable exoplanetary systems around pulsars by capture: An exercise in computational classical mechanics}

\author{%
Václav Pavlík$^{1,2,*}$\orcid{0000-0002-3031-062X},
Steven N.~Shore$^{3,4}$\orcid{0000-0003-1677-8004},
Vladimír Karas$^{1}$\orcid{0000-0002-5760-0459},
and Matyáš Fuksa$^{5}$}

\affil{$^1$ Astronomical Institute of the Czech Academy of Sciences, Boční~II~1401, 141~00~Prague~4, Czech Republic}

\affil{$^2$ Indiana University, Department of Astronomy, Swain Hall West, 727 E 3$^\text{rd}$ Street, Bloomington, IN 47405, USA}

\affil{$^3$ Dipartimento di Fisica, Universit\`a di Pisa, largo B.~Pontecorvo 3, Pisa~56127,~Italy}

\affil{$^4$ INAF-OATS, Via G.~B.~Tiepolo 11, 34143 Trieste, Italy}

\affil{$^5$ Astronomical Institute, Charles University, V~Holešovičkách~2, 180~00~Prague~8, Czech Republic}

\affil{$^*$ Corresponding author.}

\email{pavlik@asu.cas.cz}


\begin{abstract} 
The study of our Solar System---its formation, evolution, and long-term stability---has been ongoing for centuries and is now a standard part of scientific education. While the formation of other Solar-like exoplanetary systems is generally explained using the same mechanisms that describe our own, the discovery of exoplanets around pulsars in 1990s has raised new questions about their origin. Several scenarios were proposed, including formation by capture during a close encounter of a compact stellar-mass remnant and a pre-existing planetary system. It was, however, also conjectured that captured planets should exhibit high eccentricities and---if more planets are captured---their evolution would lead to chaos.   We revisit classical mechanics as applied to planetary systems. As an example and follow-up to previous works, we use an open-source high-precision $N$-body code to investigate dynamical interactions between planetary systems and stellar remnants, the orbital properties of captured planets, and their long-term stability over gigayears.
We corroborate that the captured planets often exhibit high eccentricities (unlike some observed pulsar planetary systems), but we also present a student's simulation where a Jupiter-like planet undergoes a series of planet--planet encounters and planetary ejections, eventually stabilising at a low eccentricity of ${\sim}0.146$.
This shows that a chaotic post-capture evolution may eventually lead to long-term stability, making the dynamical formation channel viable for producing low-eccentricity systems. These results warrant more detailed investigation in future work. Beyond their astrophysical significance, they also illustrate general principles of non-linear dynamics and computation, where aspects of the analysis can even be carried out at the high-school or undergraduate level, making this type of research accessible to students at an early stage.
\end{abstract}

\section{Introduction}
\label{sec:intro}

Classical mechanics, while essential in the training of a physicist, is often seen by students as a formality in anticipation of the exciting stuff, such as quantum and statistical mechanics.  Much of the current material taught in these basic courses is designed to introduce Lagrangians, Hamiltonians, and potential fields that will be needed for generalised field theory courses \citep[e.g.,][]{goldstein2002classical, jackson2021classical}.  In so doing, the fundamental origin problem of Newtonian mechanics, the universality of gravitation, is often relegated to a mere exercise in force balance or the solution of the two-body problem in a formal way.  However, since the 19th century, a core problem in mechanics has remained the study of non-linear, complex systems with multiple degrees of freedom, the most famous of which is the three-body problem of celestial dynamics.\!\footnote{This has even entered popular culture with the novel and a streaming-platform series of the same name. While the premise may be absurd, the core idea of a chaotically orbiting planet around a triple star system is well described.}

Planetary system dynamics naturally emerges as a continuation of this classical problem. It concerns the gravitational evolution of systems composed of many interacting bodies, where nonlinearity, resonances, and chaos are not mere mathematical curiosities but some of the defining features \citep{tremaine2023dynamics}. Since the time of Kepler,\!\footnote{For a historical aside unrelated to celestial dynamics, the reader may also wish to consult the discussion of Johannes Kepler’s portraits and their cultural context in \citet{kepler_portrait, kepler_portrait_ext}.} Newton, Laplace, and Poincar{\'e}, celestial mechanics has sought to understand how planetary systems form, evolve, and remain stable over astronomical timescales \citep[see, e.g.][]{tamayo+2020} or the structure of satellite systems and planetary rings \citep{1983AJ.....88.1560B, planetary_rings}.

In the last few decades, the development of reliable, publicly available numerical codes to calculate orbital elements renders much of the traditional celestial mechanics course irrelevant. But it is through celestial mechanics that the fundamental concepts of how periodic systems behave were historically introduced \citep{sommerfeld1923, sternberg1969}---for instance, perturbation theories, asymptotic series, symplectic manifolds, and stability. While these are treated in mechanics courses, they are frequently divorced from their origin. An opportunity is lost to show how methods developed in one area can be adapted to a wide range of problems, and not just in classical mechanics. In the age of commercial space development, understanding orbital mechanics is no longer a luxury.

Current advances in computational methods and dynamical systems theory---combined with the discovery of Earth-like planets around Solar-like stars and a vast diversity of exoplanetary architectures---have significantly expanded this field \citep[see, e.g.][]{bolmont+2014, valencia+2025}. The Solar System provides a benchmark for studying long-term stability and secular interactions, while extrasolar systems reveal migration histories and configurations that sometimes make us question the classical formation scenarios.

As a prime example, the origin of pulsar planets is still debated. With only three confirmed systems---PSR~1257+12 \citep{wol_fra92,wol94,gkw05}, PSR~1829--10 \citep{bls91}, and PSR~B1620--26 \citep{bfs93,arz_etal96,tho_etal99,ford_etal2000}---no firm conclusions can be drawn. One possibility is that they are relics of primordial systems that survived the red-giant phase and supernova explosion, however, this is considered improbable \citep{lwb91}. An \textit{in situ} formation of planets in a post-supernova accretion disk is more likely \citep{ppr91,phi_han93,cur_han07}, but only one such disk has been observed to date \citep{wck06}, and its reported mass is too low to allow for the birth of planets.

In a recent exploration of a scenario in which a compact massive interloper---a stellar-mass black hole or neutron star (NS)---encounters a planetary system, one exit channel resulted in planetary captures by the NS \citep[][hereafter \citetalias{death_star}]{death_star}. Although dynamical formation has already been suggested by \citet{ppr91} and proposed to explain, for instance, the origin of PSR~B1620--26 \citep{pod95, ford_etal2000}, the long-term stability of such systems has not yet been studied. Moreover, the main hypothesis presented by \citetalias{death_star} was that the capture of multiple planets would lead to a long-term chaotic behaviour.

This article presents a summary of a numerical experiment performed by an MSc student \citep{fuksaMSc} as a follow-up to \citetalias{death_star}. We aimed to showcase the accessible and straightforward nature\footnote{Accessibility in this context does not imply conceptual simplicity, but rather that the problem can be explored with (i) open-source tools, (ii) modest computational resources, and (iii) incremental conceptual scaffolding.} of this research problem (described in Section~\ref{sec:models}) that ended up providing more insights into the evolution of the captured systems on gigayear time scales.
Time constraints did not allow us to focus on the statistical aspect of the planetary captures in the long term (e.g.\ several methods are described in Section~\ref{sec:discussion}).
Instead, we present an in-depth case study of the long-term evolution of one random set of initial conditions, which we found particularly interesting (see Section~\ref{sec:results}).

\begin{figure*}
    \centering
    \includegraphics[width=.99\linewidth]{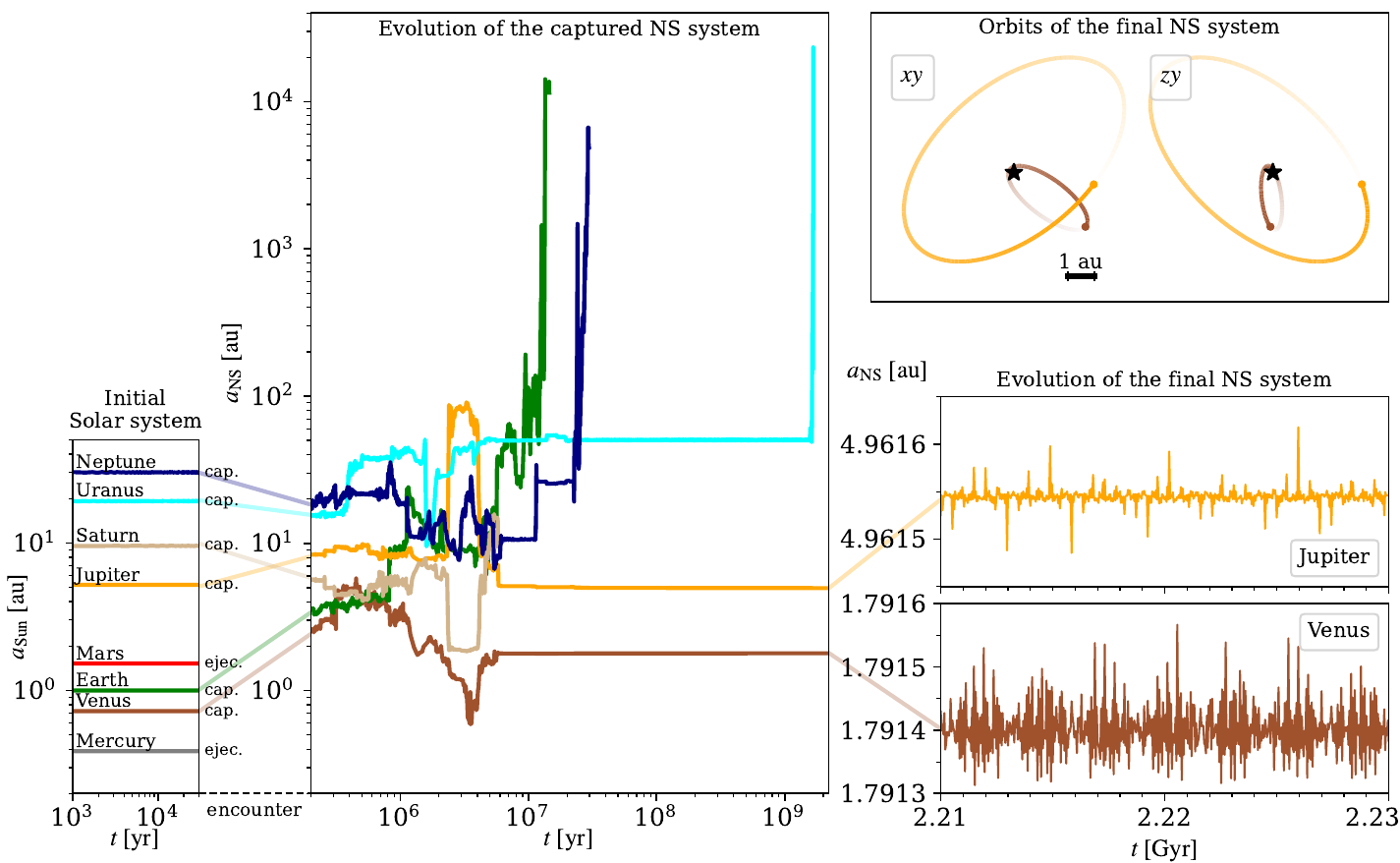}\\
    
    \caption{Evolution of the planetary semi-major axes in the model time (time zero would correspond to the start of the simulation).
    Left: Semi-major axes of the planets in the system, as the initial model set-up (both axes are in log-scale).
    Centre: Semi-major axes of the planets following the NS capture (both axes are in log-scale). We note that the encounter phase was chaotic and the calculation of the semi-major axis was not well-defined because of the temporary presence of two massive bodies in the centre of the system; therefore, this plot continues after we have removed the Sun from the simulation.
    Bottom right: Detail of the semi-major axes of Jupiter and Venus in the final NS system (both axes are in linear scale and the time range in Gyr corresponds to the detail in Fig.~\ref{fig:ecc}).
    Top right: Orbits of the final system at the end of the simulation in two different projections.}
    \label{fig:semi}
\vspace{\floatsep}
    \centering
    \includegraphics[width=.99\linewidth]{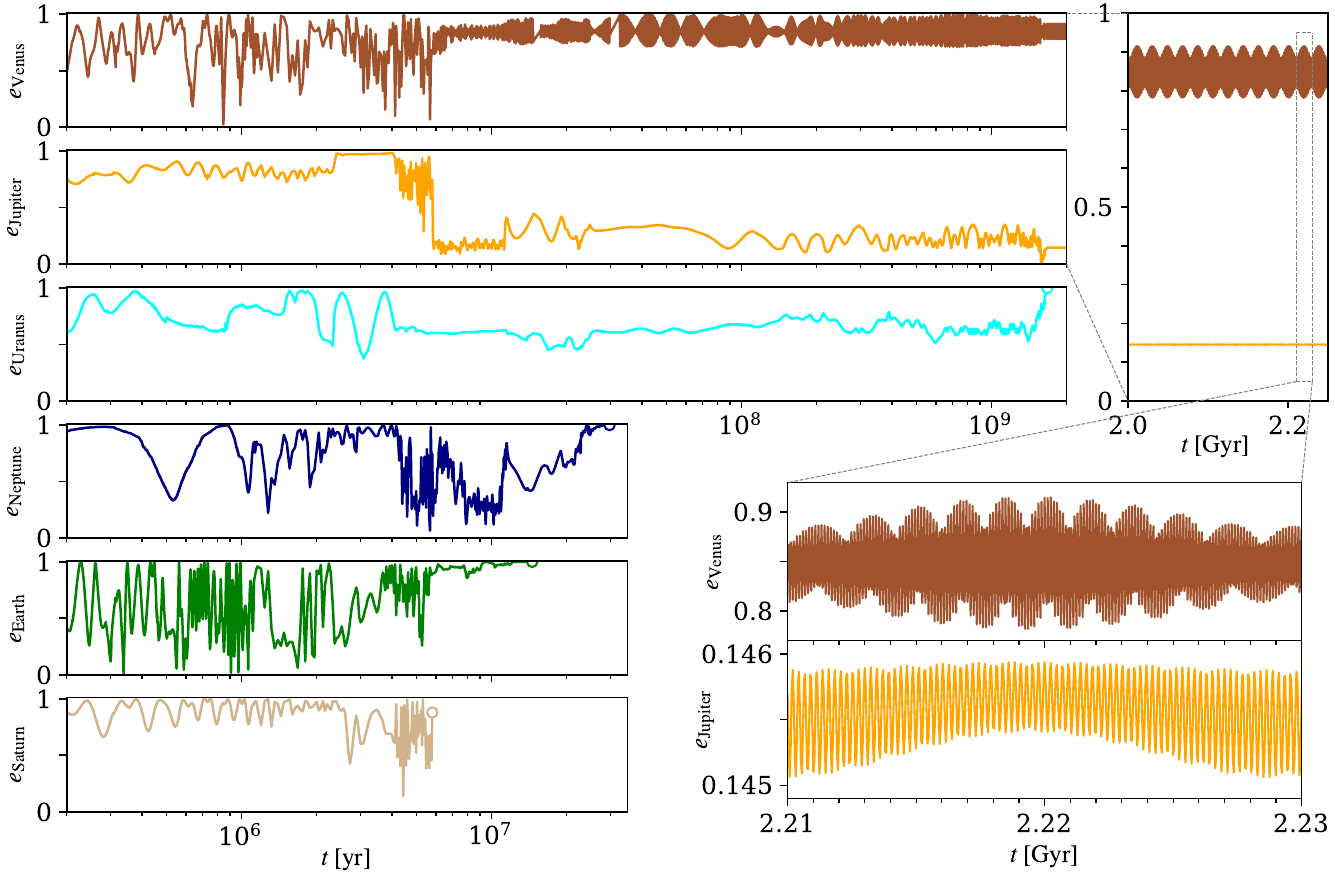}
    
    \caption{Eccentricities of the captured planets (circles mark escapes from the NS system).
    The left-hand panels show the long-term evolution, where the horizontal axis is the model time in years on a logarithmic scale.
    The two zoomed-in right-hand panels show the final system with Jupiter and Venus in gigayears with time on a linear scale (and the time range of the bottom-right plot corresponds to the detail in Fig.~\ref{fig:semi}). The colours are the same as in Fig.~\ref{fig:semi}.}
    \label{fig:ecc}
\end{figure*}

\section{Numerical setup}
\label{sec:models}

Our adopted initial set-up is similar to \citetalias{death_star}.
We simulated encounter scenarios between the Solar System\footnote{Parameters from NASA JPL HORIZONS as of 01/01/2000.} (as a test-bed for a known and stable planetary system; see, e.g., the left-hand panel of Fig.~\ref{fig:semi}) and a two-solar-mass body (hereafter NS).
All bodies were treated as point-masses, and the evolution was tracked with the \textsc{IAS15} integrator from the \textsc{Rebound} package \citep{rebound}, which offers high accuracy and reproducibility \citep{reboundsa}, and is open source and community supported.

A brief comment on the numerical approach is warranted. The \textsc{Rebound} package is a general-purpose $N$-body integrator specifically developed for planetary system dynamics, offering a suite of integration schemes---including the classical leapfrog, symplectic methods such as \textsc{WHFast} \citep{wisdom_holman1991, WHFast}, as well as the high-order, adaptive, non-symplectic integrator \textsc{IAS15} \citep{everhart1985, IAS15} employed in this work.
Its principal strength lies in the ability to accurately follow the long-term evolution of systems consisting of a modest number of gravitating bodies (i.e.\ up to $N \sim 10^2{-}10^3$), where the conservation of orbital elements and faithful treatment of secular interactions or resonances are essential.
Since the code relies on direct summation of gravitational forces, it implies an $\mathcal{O}(N^2)$ scaling with particle number.
While this renders \textsc{Rebound} inefficient for large-$N$ problems such as dense star clusters ($N \sim 10^4{-}10^6$) or galactic nuclei ($N \sim 10^7$), this software is particularly relevant for planetary systems evolving over millions to billions of years, as considered in our work.\!\footnote{We have also recently adopted \textsc{Rebound} to follow the dynamical stability of less populous stellar associations in the vicinity of the super-massive black hole in the centre of our Galaxy since the scaling of the central body to the stars was similar to the ratio of masses between the Sun and its planets \citep{pavl_irs13}.}$^,$\footnote{And as a popular-culture aside, the same class of direct $N$-body tools could equally be used to model hypothetical planetary orbits around compact objects such as supermassive black holes (e.g.\ as depicted in the film \textit{Interstellar}). While the astrophysical realism of such scenarios depends sensitively on relativistic effects not included here, the underlying Newtonian dynamics and numerical methodology are conceptually identical. To add relativistic corrections, we refer the reader to the \textsc{ReboundX} library \citep{reboundx}.}

Its modular, open-source design further allows for flexible choices of integrators, collision handling, and boundary conditions, making it a standard tool in contemporary computational celestial mechanics and an ideal tool for teaching students how to perform numerical simulations.

In our numerical experiments, we varied the NS's impact parameter, heliocentric velocity, and incidence angle with respect to the ecliptic, reproducing the initial conditions from \citetalias{death_star}.  The novelty here is, instead, that knowing the outcomes of the NS encounters from \citetalias{death_star}, we chose those initial conditions for which the capture of more than one planet has been reported (see also Appendix~\ref{app:models} for the list of physical initial conditions).  We followed these systems for additional millions to billions of years in simulation time, which is far longer than what was performed in \citetalias{death_star}, to assess the long-term dynamical stability of the captured planetary systems. All escaping bodies were removed when their distance from the central body reached 1\,pc to optimise the computations. Those escapers comprise all uncaptured planets and the Sun after the initial encounter, and the captured planets that are subsequently ejected from the system after mutual close encounters.

\section{An example of a low-eccentricity system}
\label{sec:results}

The full range of initial conditions we tested and their preliminary discussion is in \citetalias{death_star} and \citet[][see also Appendix~\ref{app:models}]{fuksaMSc}.
In this section, we focus on one particular case of captured planets that initiate in a chaotic state and pass to stability as a low-eccentricity planetary system.
In this scenario, the NS was initially launched with the heliocentric velocity of $v_\infty = 10\,\kms$, from $i=30^\circ$ ecliptic latitude, and almost head-on with $b=0.2\,\au$ impact parameter.

This is a very close encounter, which is why the capture of multiple planets --- rather than the destruction of the entire Solar System --- is possible. With gravitational focusing, the effective cross section for encounters closer than $b$ for bodies of masses $m_1$ and $m_2$ is
\begin{equation}
    \label{eq:cross_section}
    \sigma = \pi b^2 \left[ 1 + \frac{2 G (m_1 + m_2)}{b v_\infty^2} \right] \,,
\end{equation}
where $G$ is the gravitational constant \citep[see also, e.g.,][]{binney_tremaine}. In an environment of stellar density $n$, the encounter rate per star is
\begin{equation}
    \label{eq:enc_rate}
    R = n \sigma v_\infty \,.
\end{equation}
For the masses, velocity, and impact parameter from our simulation ($m_1 = 1\,\Msun$\,, $m_2 = 2\,\Msun$, $v_\infty = 10\,\kms$, $b=0.2\,\au$), we get
    \hbox{$R \approx 8.1 \cdot 10^{-9} \, (n / \pc^{-3}) \, \Myr^{-1}$}.
For completeness, we note that the mean encounter rate per star that could lead to capture for any impact parameter and incidence angle has already been estimated in \citetalias{death_star} as
    \hbox{$3\times10^{-7}\,(n/\pc^{-3})\,\Myr^{-1}$}.
Therefore, several captured systems could form, e.g.\ in a globular cluster over its lifetime.

After the encounter with the Solar System, the NS in our simulation captured six planets on highly eccentric orbits (see the left-hand panels of Fig.~\ref{fig:ecc}). The captured system subsequently underwent substantial changes, which we describe in the following.

The interactions between Saturn and Jupiter on eccentric orbits ($e_\mathrm{S} \approx e_\mathrm{J} \approx 0.8$ on average) caused large variations in their semi-major axes; see the yellow and beige lines in the central panel of Fig.~\ref{fig:semi} where they are in anti-phase. This caused rapid, large fluctuations in eccentricities of both terrestrial planets (see the panels of Earth and Venus in Fig.~\ref{fig:ecc}) and also somewhat longer-period oscillations in the eccentricity of the other planets in the system.

Several close encounters between Jupiter and Saturn at $t \approx 2.2\,\Myr$ sent the former to an extremely elongated trajectory with the semi-major axis $a_\mathrm{J} \approx 100\,\au$, and the latter to an orbit with $a_\mathrm{S} \approx 2\,\au$ (see Figs.~\ref{fig:semi} and~\ref{fig:ecc}).
We found a subsequent mutual exchange of angular momentum between Saturn, Uranus, Earth, and Neptune (mainly observed in the variations of eccentricity) which lasted until $t \approx 4\,\Myr$. Then another series of close encounters between Jupiter and Saturn began in the central regions of the system. This resulted in large fluctuations of Saturn's eccentricity and led to its ejection at $t \approx 6\,\Myr$. The removal of this second-most massive planet lowered Jupiter's eccentricity significantly, and injected both Venus and Earth into very elongated orbits (see Fig.~\ref{fig:ecc}). But it also stabilised the semi-major axes of Jupiter, Venus, and---to some degree---also that of Uranus (see Fig.~\ref{fig:semi} after $t \approx 6\,\Myr$).

After $t \approx 12\,\Myr$, we found strong ongoing interactions of Neptune, Jupiter, and Uranus that destabilised the Earth's orbit and led to its ejection at $t \approx 15\,\Myr$, followed by the unbinding of Neptune at $t \approx 30\,\Myr$. The three-planet system evolved in a semi-stable way for more than a Gyr with small variations in their eccentricities and almost no variations in the semi-major axes. But it eventually led to the last close encounter between Jupiter and Uranus at $t \approx 1.6\,\Gyr$, which ejected the latter and stabilised Jupiter's eccentricity at a low value of $e_\mathrm{J} \approx 0.1455$ (see Fig.~\ref{fig:ecc}). This produced the final NS system consisting of the gas giant and Venus (with $e_\mathrm{V} \approx 0.85$) in non-coplanar orbits (mean mutual inclination $i_\mathrm{V}-i_\mathrm{J} \approx 60^\circ$), see the illustrative orbits in the top right corner of Fig.~\ref{fig:semi}.

The bottom-right panels of Fig.~\ref{fig:semi} show that both planets are on stable orbits with small fluctuations around the mean values of the semi-major axes $a_\mathrm{J} \approx 4.96155\,\au$ and $a_\mathrm{V} \approx 1.7914\,\au$.
Specifically, the most noticeable period of fluctuations in Venus' orbit is $T_{a} \approx 2.8\,\Myr$ (and we note that the tallest downward spikes in $a_\mathrm{J}$ appear to coincide with the regions of minimum variations of $a_\mathrm{V}$).

The right-hand panels of Fig.~\ref{fig:ecc} further show that Venus and Jupiter periodically exchange angular momentum, which affects the evolution of their orbital eccentricities. Specifically, the bottom-right-hand panel reveals three periods. The longest variations have $T_{e1} \approx 19\,\Myr$ and the fastest ones \hbox{$T_{e3} \approx 0.12\,\Myr$}. The intermediate oscillations span a range of periods \hbox{$T_{e2} \in (1.7;2.6)\,\Myr$}, where the lower limit corresponds to the largest change of $a_\mathrm{V}$ and the upper limit (which is comparable to $T_a$) to the smallest change of $a_\mathrm{V}$ (vice versa with $a_\mathrm{J}$).

Since this final system has two orbiting bodies---lower-mass Venus inside the orbit of more massive Jupiter---we also tested whether Venus' resonances resemble those in the classical Kozai--Lidov problem \citep[hereafter KL; see][]{kozai,lidov}, or the eccentric KL \citep[hereafter eKL;][]{lithwick_naoz}, taking into account the non-zero eccentricity of Jupiter. We found no clear evidence for either. The main issue is that both KL and eKL assume that the gravitational effects from the perturbed body (i.e., Venus) on the perturbing body (i.e., Jupiter) are negligible and that the Hamiltonian of the perturbed body is separately conserved \citep{poisson_will}. Neither is the case in this system due to the proximity of both planets in the final system ($a_\mathrm{J} - a_\mathrm{V} \approx 3.2\,\au$), and their comparatively larger mass ratio ($M_\mathrm{V}/M_\mathrm{J} \approx 2.6\times10^{-3}$). Hence, the Hamiltonian of Venus' orbit also changes.

We integrated this final NS system for several {\it gigayears}  beyond what is shown in the figures, and we did not observe any notable change in the planets' behaviour.  Therefore, the final NS planetary system appears stable on a time scale comparable to the age of galaxies.

\section{Discussion and implications for teaching}
\label{sec:discussion}

\subsection{Celestial mechanics and modern classical physics}

Celestial mechanics has usually been relegated to specialist mathematics courses and has almost completely disappeared from the standard curriculum.  If mentioned at all in classical mechanics courses, it is only to show closed-form solutions to the Kepler problem.  Yet there are remnants of the subject lurking in the standard texts \citep[e.g.][]{goldstein2002classical} in the discussion of canonical transformations.

In the development of quantum mechanics, notwithstanding the Bohr atom, canonical variables and the Hamiltonian (and Hamilton--Jacobi equation) were central to the formalism \citep{shore2003, fraser_nakane2023}.  This was because of the celestial mechanical roots of the transformations in passing from Cartesian to specialised coordinates (e.g.\ the Delaunay variables) that provide evolution equations for many-body systems.  These are embodied in the symplectic integrators used in few-body integrators such as \textsc{Rebound}.  Consequently, one of the most abstract subjects in the current presentation of classical physics can be rendered more comprehensible through the sort of simulation we have presented here. From an educational perspective, this conceptual reframing also provides a natural entry point for introducing students to modern computational approaches to classical mechanics.

\begin{figure}[t]
    \centering
    \input{flowchart}
    \caption{Conceptual pathway illustrating how the two-body problem develops into few-body dynamics, chaos, and statistical mechanics. It also highlights the parallel roles of theoretical insight and numerical methods in understanding complex systems.}
    \label{fig:flowchart}
\end{figure}
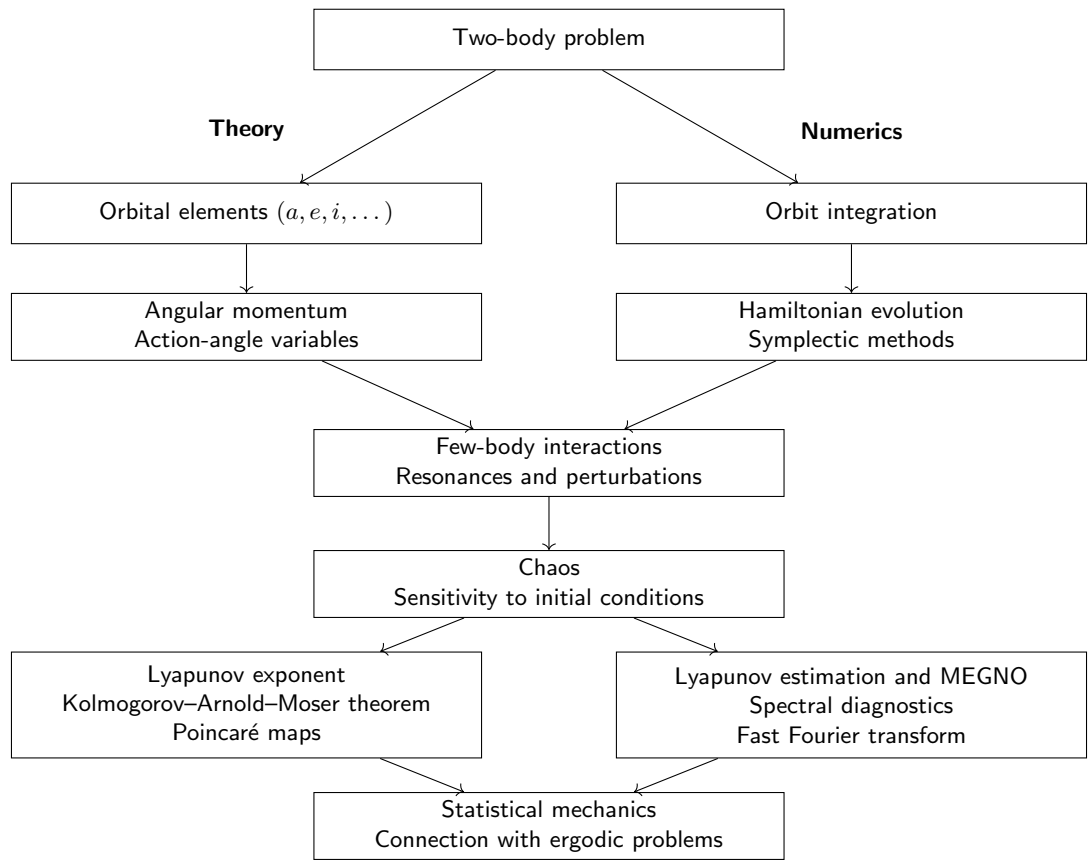

\subsection{Chaos, predictability and limits of modelling}

Celestial mechanical problems are purely conservative Hamiltonian systems and were the original context for what became known as deterministic indeterminacy or chaos \citep{poincare1892, birkhoff1966, arnold1978}.
Numerical simulations of planetary and stellar systems are fundamentally constrained by their extreme sensitivity to initial conditions---a defining feature of chaotic dynamical systems. Even infinitesimal differences in the initial phase space (i.e.\ in displacements or velocities) may grow exponentially and produce qualitatively different orbital histories at later times. This is even though the governing equations of motion are deterministic.
This exponential sensitivity to the orbital elements is often quantified by the Lyapunov exponent, which measures the rate at which nearby trajectories diverge in phase space, and whose inverse (the Lyapunov time) gives an estimate of the horizon beyond which detailed prediction of individual orbits is effectively impossible \citep{lyapunov1966, benettin_galgani1979, cincotta_simo2000, cencini_ginelli2013}. A positive maximal Lyapunov exponent thus signals chaos, and small observational or numerical uncertainties amplify rapidly, rendering precise orbital predictions near impossible on much longer timescales \citep{hussain_tamayo2019}. This transition from integrable behaviour to chaos and its diagnostic characterisation is illustrated schematically in Fig.~\ref{fig:flowchart}.

For instance, the Solar System's Lyapunov time is of order millions of years (commonly quoted $\tL \sim 5\,\Myr$ for the inner terrestrial planets and $\tL \sim 10{-}20\,\Myr$ for the outer gas giants). The Solar System is chaotic yet likely stable over gigayear timescales, with very low but non-zero probabilities of extreme events (e.g.\ Mercury destabilisation) \citep{mogavero+2023}.

From an educational perspective, this provides a concrete setting in which students at any level can confront a key epistemic insight that deterministic equations do not imply predictable outcomes over long timescales. At the high-school level, this can be approached qualitatively by comparing two nearly identical simulations (or even precomputed trajectories) and observing how small differences in initial conditions lead to visibly different orbital evolutions. This would introduce the idea of sensitivity to initial conditions even without formal mathematical machinery. At the undergraduate or graduate level, students can actively implement such experiments using a numerical tool (e.g.~REBOUND) to initialise two systems with a small velocity perturbation (e.g.\ $\Delta v/v \lesssim 10^{-6}$) and to track the divergence of orbital elements (e.g.~semi-major axis or eccentricity). This will allow them to gain hands-on experience with numerical integration and the limits of predictability and to extend to a more quantitative analysis, including estimating the rate of divergence and relating it, for instance, to the concept of the Lyapunov exponent (discussed above).

Because of this fundamental limit, the interpretation of long-term integrations must usually move away from the expectation of unique predictions and toward a statistical and diagnostic approach. Practically, this means characterising ensembles of evolutions, mapping instability timescales, and employing chaos indicators---e.g.\ MEGNO (Mean Exponential Growth factor of Nearby Orbits) and other fast chaos detectors \citep{skokos2010}---and spectral diagnostics (e.g. with the Fast Fourier transform) to distinguish regular from chaotic behaviour in numerical outputs \citep{gemmeke+2006}.
Furthermore, Kolmogorov--Arnold--Moser theory \citep[KAM;][]{Kolmogorov1954, Arnold1963, Moser1973} provides a theoretical framework explaining why regular, quasi-periodic motion can persist alongside chaos in weakly perturbed Hamiltonian systems.
Such analyses were beyond the scope of this paper; however, we refer the interested reader to our earlier study \citepalias{death_star}.

A complementary and particularly intuitive approach to analysing such dynamical behaviour is the Poincaré return map \citep{poincare1892, fraser_nakane2023}. It is a standard analytical tool for dynamical systems that is introduced---if at all---in courses on differential equations or in engineering-related courses on feedback.  It originated in the three-body problem and Poincaré's topological approach.  The concept is easily understood even without advanced analysis since it is a mapping of the orbit onto a projected plane in a phase space.  Even if the student has not seen statistical mechanics, this is an ideal way of visualising an autonomous system since it deals with orbits, for which phase and time are interchangeable.  It is also an excellent way to visualize the approach to chaos and can be directly compared between simulations with different orbital parameters and numbers of bodies.  Finally, figuring out how to write the algorithm as an automated procedure---as in this project---is an instructive application of post-processing of the models \citep[see also][]{Minorsky1962}.

It is also worth mentioning that in parallel, data-driven and machine-learning methods have recently been developed to predict stability boundaries and to classify the long-term behaviour of compact multi-planet systems \citep{tamayo+2020}. This shift is especially important for exoplanet studies where observational uncertainties in orbital parameters can place apparently acceptable orbital fits on chaotic trajectories that rapidly evolve toward collisions or ejections. Conversely, configurations that appear unstable when judged from observational fits alone may persist over much longer timescales, nonetheless. Stability analysis, therefore, provides an important tool for assessing and constraining the plausible architectures of multi-planet systems \citep{heggie1991,gajdos_vanko2023}.

\subsection{Conceptual links and broader relevance}

From a dynamical point of view, planetary capture during a close flyby can be viewed as a gravitational analogue of classical scattering. Similarly to Rutherford scattering---which is best known from particle physics or chemistry courses---the outcome of the encounter is largely controlled by the asymptotic velocity and impact parameter. These determine the strength and duration of the interaction. Gravitational focusing then defines an effective capture radius and enhances energy exchange during slow, close encounters \citep[as emphasised in the theory of stellar encounters by][]{Chandrasekhar1943}. In the language of few-body dynamics, such interactions resemble the exchange reactions described by the Hills--Heggie law, i.e.\ that loosely bound systems are disrupted while more tightly bound ones preferentially survive \citep{Hills1975, Heggie1975}. Planetary capture may, therefore, be interpreted as a low-mass analogue of these processes, with planets being ejected, exchanged, or retained depending on the encounter geometry. Similarly, when focused on an even lower-mass scale, close stellar encounters may cause disruptions of the orbits of small bodies (e.g.\ cometary nuclei) within the system \citep[e.g.][]{oort_cloud, gliese_710}, or their unbinding, leading to the production of interstellar asteroids and comets \citep{1I_Oumuamua, 2I_Borisov, 3I_Atlas}.\!\footnote{The discovery---to date---of three interstellar comets presents the inverse capture problem.  The particular interest is that the dynamics are regularly updated during passage and available in real time, so these elements can be used to study orbital prediction and {\it retrodiction}.}

In a broader context, dynamical astronomy also offers a natural entry point for student engagement at the advanced secondary and early undergraduate levels. Current research problems---ranging from stellar dynamics in galactic nuclei to planetary encounters and capture---can be explored using publicly available data, open-access literature, and well-documented numerical tools. These allow students to work with the same conceptual framework as professional researchers.\!\footnote{For example, international and national Astronomy Olympiads---extracurricular competitions held worldwide and supported by Division~C of the International Astronomical Union---provide early engagement with astronomy, support the development of reasoning skills in students, and serve as a natural entry point for further involvement in research \citep{pavlik_ao2024}.}
Through such projects, students are introduced not only to Keplerian motion and gravitational interactions, but also to tidal effects, accretion processes, and the dynamical environment of galactic centres hosting supermassive black holes. Moreover, early exposure to research helps students learn how to navigate scientific literature and to critically assess scientific problems. Consequently, classical mechanics and dynamical astronomy offer an effective foundation for developing transferable skills and physical intuition early in academic training.

\section{Conclusions}
\label{sec:concl}

Studying the dynamics of captured planetary systems advances our understanding of pulsar planets and provides a natural context for student engagement. By working with $N$-body simulations and chaos diagnostics, students can explore real research problems and develop physical intuition for orbital dynamics and the long-term evolution of astrophysical systems.
In this context, the present study also reflects the structure of an MSc-level project, which combined a well-defined physical framework with an open-ended outcome. The availability of robust numerical tools and accessible initial conditions made it possible to focus on interpretation rather than technical implementation. At the same time, the intrinsic sensitivity of the system required iterative reasoning and critical assessment of results, which are central aspects of research practice.

We showed that the long-term evolution of captured systems with multiple planets may go through a temporary chaotic evolution and eventually stabilise. The remaining planetary orbits decrease their initially high eccentricities through planet--planet interactions that lead to ejections of other planets.
This partially answers the question posed in \citetalias{death_star}: The captured NS planetary systems do not always end up in a chaotic state, and it is possible to produce low-eccentricity orbits through this channel.

On the other hand, we also understand the limitations of our models as well as the lack of statistics in the observed data. For instance, from the three known pulsar systems, only the planet in PSR~B1620--26 has a comparable eccentricity to Jupiter from our final simulation. However, the central NS in PSR~B1620--26 is part of a binary system, whereas our NS is not. Moreover, we sourced the planets from specific initial conditions---the Solar System---thus, the modelled capture scenario might be biased in a way.

Nonetheless, we must emphasise that the presented simulation was a random realisation of those initial conditions---we did not design it to produce a low-eccentricity planetary system. Given the chaotic nature of the post-capture dynamical evolution, trying to arrange such an outcome would have been near-impossible, regardless.
Therefore, we conclude that our case study produced viable results for the possible origin of pulsar planets and showed that planetary captures by stellar remnants are worth exploring in a wider parameter space with more diverse initial conditions and external perturbations, e.g., in a star cluster environment.

Finally, this study illustrates how active research can be fitted into a teaching curriculum. The pedagogical framing follows the structure of the problem itself. Understanding chaotic $N$-body systems relies on numerical experimentation, statistical interpretation, and the assessment of model-dependent outcomes. Planetary dynamics thus provides a hands-on way for students to develop scientific reasoning while working on current questions in astronomical research. 

%
%

\ack{
We used computational resources from e-INFRA CZ (project ID:90254), supported by the Ministry of Education, Youth and Sports of the Czech Republic.
We are also grateful we had the opportunity to present part of this research at the MODEST-24 conference \citep{fuksa_modest}.
}

\funding{
VP is funded by the European Union's Horizon Europe and the Central Bohemian Region under the Marie Skłodowska-Curie Actions -- COFUND, Grant agreement \href{https://doi.org/10.3030/101081195}{ID:101081195} (``MERIT'').
VP and VK acknowledge support from the project RVO:67985815 at the Czech Academy of Sciences.
}


\data{The data that supports the findings of this study are available from the corresponding author upon reasonable request.}


\setlength{\bibsep}{0.0pt}
\bibliographystyle{aa}
\bibliography{main}

\clearpage
\appendix

\section{Parameters of the numerical experiment}
\label{app:models}

For the reader's convenience, here we list the initial conditions of all models studied in the MSc thesis of \citet{fuksaMSc}, which is a subset of models published in \citetalias{death_star}. Both of these works have already established the initial conditions leading to the capture of planets by the massive remnant.
Specifically, out of the 156 combinations of the initial parameters listed in Table~\ref{tab:init}, sixty models resulted in the capture of at least one planet (ten of them had two or three planets, and twenty had four or more planets).
\citet{fuksaMSc} has followed their evolution for another $1$ to $2\,\Myr$, and then selected 11 simulations (see Table~\ref{tab:long}) which showed interesting properties -- e.g., high number of planets or their types, planets in mutually retrograde orbits or hint of resonances. They then tracked the evolution of these sets of initial conditions for another $1$ to $10\,\Gyr$.

The model discussed in this paper (marked in Table~\ref{tab:long}) was one of those that stabilised one of its Jupiter-size captured planets at a low-eccentricity value. It showed that a captured system, which is initially chaotic, can still dynamically stabilise within a reasonable time frame. Thus, we dedicated a more in-depth analysis to it in our paper.

\begin{table}[!h]
    \centering
    \caption{Initial conditions of the models}
    \begin{tabular}{ll}
        \hline
        \textbf{Solar System} &\\
        Initial positions and masses: & taken from NASA JPL HORIZONS as of January 1, 2000 \\
        \hline
        \textbf{Intruder} &\\
        Mass: & neutron star (NS; $m=2\,\Msun$) or black hole (BH; $m=10\,\Msun$) \\
        Distance from the Solar System: & $d \geq 0.5\,\pc$ \\
        Heliocentric velocity: & $v \in \{ 10; 50 \} \,\kms$ \\
        Impact parameter: & $b \in \{ 10^{-6}; 10^{-5}; 10^{-4} \} \,\pc$ \\
        Incidence angle from the ecliptic: & $\theta \in \{ 0^\circ; 15^\circ; 30^\circ; 45^\circ; 60^\circ; 75^\circ; 90^\circ; 105^\circ; 120^\circ; 135^\circ; 150^\circ; 165^\circ; 180^\circ \}$ \\
        \hline
    \end{tabular}
    \label{tab:init}
\end{table}

\begin{table}[!h]
    \centering
    \caption{Initial conditions and properties of the long-term simulations}
    \begin{tabular}{cccrl}
        Intruder & $v\,/\,\kms$ & $b\,/\,\pc$ & $\theta\,/\,^\circ$ & Number of captured planets after $2\,\Myr$ \\
        \hline
        NS & 10 & $10^{-6}$ & 30 & 6 planets -- the model described in this paper \\
        NS & 10 & $10^{-6}$ & 60 & 7 planets \\
        NS & 10 & $10^{-5}$ & 15 & 2 giant planets \\
        NS & 10 & $10^{-5}$ & 60 & all 8 planets  \\
        NS & 10 & $10^{-4}$ &  0 & 2 giant planets \\
        NS & 10 & $10^{-4}$ & 30 & 3 giant planets \\
        NS & 50 & $10^{-6}$ & 45 & 5 planets \\
        BH & 10 & $10^{-6}$ &  0 & 3 giant planets \\
        BH & 10 & $10^{-6}$ & 60 & all 8 planets \\
        BH & 10 & $10^{-5}$ &  0 & 2 giant planets \\
        BH & 50 & $10^{-6}$ & 60 & 3 terrestrial planets \\
        \hline
    \end{tabular}
    \label{tab:long}
\end{table}

\end{document}

%% file: flowchart.tex

\begin{tikzpicture}

\small\sf

\node[draw, rectangle, align=center, text width=6cm, minimum height=8mm] (root)
at (0,0)
{Two-body problem};

\node[align=center] (theorylabel) at (-4,-1.2) {\textbf{Theory}};
\node[align=center] (numlabel)    at (4,-1.2)  {\textbf{Numerics}};

\node[draw, rectangle, align=center, text width=6cm, minimum height=8mm]
(theory1) at (-4,-2.3)
{Orbital elements $(a,e,i,\dots)$};

\node[draw, rectangle, align=center, text width=6cm, minimum height=8mm]
(theory2) at (-4,-3.8)
{Angular momentum\\Action-angle variables};

\node[draw, rectangle, align=center, text width=6cm, minimum height=8mm]
(num1) at (4,-2.3)
{Orbit integration};

\node[draw, rectangle, align=center, text width=6cm, minimum height=8mm]
(num2) at (4,-3.8)
{Hamiltonian evolution\\Symplectic methods};

\node[draw, rectangle, align=center, text width=6cm, minimum height=8mm]
(merge) at (0,-5.6)
{Few-body interactions\\Resonances and perturbations};

\node[draw, rectangle, align=center, text width=6cm, minimum height=8mm]
(chaos) at (0,-7.2)
{Chaos\\Sensitivity to initial conditions};

\node[draw, rectangle, align=center, text width=6cm, minimum height=14mm]
(chaosL) at (-4,-8.8)
{Lyapunov exponent\\Kolmogorov--Arnold--Moser theorem\\Poincaré maps};

\node[draw, rectangle, align=center, text width=6cm, minimum height=14mm]
(chaosR) at (4,-8.8)
{Lyapunov estimation and MEGNO\\Spectral diagnostics\\Fast Fourier transform};

\node[draw, rectangle, align=center, text width=6cm, minimum height=8mm]
(final) at (0,-10.4)
{Statistical mechanics\\Connection with ergodic problems};

\draw[->] (root) -- (theory1);
\draw[->] (root) -- (num1);

\draw[->] (theory1) -- (theory2);
\draw[->] (num1) -- (num2);

\draw[->] (theory2) -- (merge);
\draw[->] (num2) -- (merge);

\draw[->] (merge) -- (chaos);

\draw[->] (chaos) -- (chaosL);
\draw[->] (chaos) -- (chaosR);

\draw[->] (chaosL) -- (final);
\draw[->] (chaosR) -- (final);

\end{tikzpicture}